\documentclass{iaus}
\usepackage{graphicx}

\title[Formation of spheroids by mergers] 
{Formation and evolution of galactic spheroids by mergers}

\author[Naab et al.]   
{Thorsten Naab$^1$%
  \thanks{email: naab@usm.lmu.de},
 Andreas Burkert$^1$, Peter H. Johansson$^1$, \break \and Roland Jesseit$^1$}

\affiliation{$^1$University Observatory Munich,
Scheinerstr. 1, 81679 Munich, Germany \break email:naab@usm.lmu}

\pubyear{2004}
\volume{xxx}  
\pagerange{119--126}
\date{?? and in revised form ??}
\jname{Proceedings Title IAU Symposium}
\editors{A.C. Editor, B.D. Editor \& C.E. Editor, eds.}
\begin{document}

\maketitle

\begin{abstract}
Galactic spheroids can form as a result of galaxy interactions and
mergers of disks. Detailed analyses of the photometric properties, the
intrinsic orbital structure, the line-of-sight velocity distributions and
the kinemetry of simulated merger remnants, which depend critically on the
geometry and the gas content of the interacting progenitors, indicate that 
low and intermediate mass rotating ellipticals can form from mergers of
disks. The masses and metallicities of all massive ellipticals and the 
kinematics of some massive non-rotating ellipticals cannot be
explained by binary mergers. Thus these galaxies might have 
formed in a different way.  

\end{abstract}

\firstsection 
\section{The disk merger hypothesis} 
Following the first fully self-consistent simulations of mergers of
stellar disk galaxies performed by \cite[Gerhard (1981) and Negroponte \&
  White (1983)]{1981MNRAS.197..179G,1983MNRAS.205.1009N} a large number
of increasingly sophisticated simulations have been performed to test
whether disk mergers can form elliptical galaxies (see
e.g. \cite[Barnes \& Hernquist 1992]{1992ARA&A..30..705B}).
After more than three decades of research we can summarize the
situation as follows: global properties of disk merger remnants are in
several respects consistent with observations of giant elliptical
galaxies, e.g. equal mass remnants are triaxial and slowly rotating,
have boxy or disky isophotes \cite[(Heyl et al. 1994, Naab, Hernquist
  \& Burkert 1999, Naab \& Burkert 2003)]{1999ApJ...523L.133N,1994ApJ...427..165H,2003ApJ...597..893N}. In
addition, mergers of disks can result in the formation of kinematic
subsystems like kinematically decoupled cores at the centers of
ellipticals \cite[(Hernquist \& Barnes 1991, Jesseit et
  al. 2007)]{1991Natur.354..210H,2007MNRAS.376..997J} as well as 
observed faint structures like shells, loops and ripples at large
radii. Unequal mass mergers are more supported by rotation and have
disky isophotes \cite[(Naab \& Burkert
  2003, Bournaud et al. 2005)]{2005A&A...437...69B,2003ApJ...597..893N}. Collisionless merger remnants are
dominated by box orbits at their centers and tube orbits in the
outer parts. The total fraction of tube to box orbits increases  with the
mass ratio of the mergers \cite[(Jesseit et
  al. 2005)]{2005MNRAS.360.1185J} and it also correlates 
with the shape and kinematics of the systems. In addition the mix of disky and
boxy isophotal shapes for equal-mass remnants can be understood by the
projected properties of tube orbits in triaxial potentials 
\cite[(Jesseit et al. 2005, Naab et
  al. 2006a)]{2005MNRAS.360.1185J,2006MNRAS.372..839N}. It has been
  shown by \cite{1996ApJ...471..115B} that gas accumulating at the center of 
merger remnants creates a steep cusp in the central potential well
resulting in a more axisymmetric central shape of the remnants. At the
same time the fraction of stars on box orbits is significantly reduced
and tubes become the dominant orbit family. The most reasonable 
explanation for this behaviour is that systems with steep cusps in
their potential cannot sustain a large population of box orbits.
One interesting consequence of the change of orbit populations is that
the asymmetry of the LOSVD of the stars changes in a way that is
consistent with observations  making it a good dynamical tracer for
the presence of gas during the merger \cite[(Naab et
  al. 2006a)]{2006MNRAS.372..839N}.  In addition, this asymmetry can
also be thought of as a consequence of two distinct components, e.g. a
hot spheroidal bulge and a rotationally supported cold disk as
investigated by \cite{2001ApJ...555L..91N}. 

\begin{figure}
\begin{center}
 \includegraphics[width=12cm]{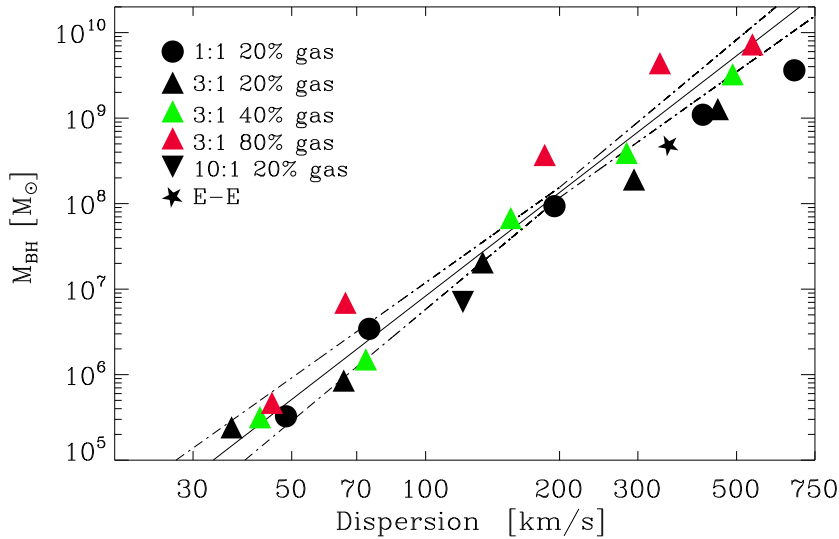}
  \caption{The final stellar velocity dispersion within the effective radius
  plotted against the final black hole mass for our sample of 1:1, 3:1, 10:1
  and E-E mergers. We overplot our simulated data with the \cite{2002ApJ...574..740T} 
  fit (solid line) to the observations together with
  their fitted 1 $\sigma$ limits (dashed lines). The simulated merger remnants follow
  the observed $M_{\rm BH}-\sigma$ relation with the scatter primarily caused
  by the initial gas mass fraction.}\label{naab_fig1}
\end{center}
\end{figure}

\cite{2003ApJ...597..893N} and \cite{2006MNRAS.369..625N} have argued,
based on statistics of kinematic and photometric properties of equal-
and unequal-mass mergers, that disk mergers (with bulges) can result
in low and intermediate mass elliptical galaxies, as seen in local observed
disk mergers \cite[(Dasyra et al. 2006a, 2006b, V\"ais\"anen et
  al. 2007)]{2006ApJ...638..745D,2006ApJ...651..835D,2007arXiv0708.2365V}. But the
objects formed in the simulations are not in agreement with the most massive, boxy and
slowly rotating  ellipticals which might have assembled by
collisionless mergers of early-type galaxies as indicated by  
observational  as well as theoretical evidence (\cite[Khochfar \&
  Burkert 2003, 2005, Khochfar \&
  Silk
  2006, Naab et al. 2006b, Bell et al. 2006]{2003ApJ...597L.117K,2005MNRAS.359.1379K,2006MNRAS.370..902K,2006ApJ...636L..81N,2006ApJ...640..241B}). 

Dissipational merging, including star formation, can also overcome
stellar phase space constraints and therefore the a priori inclusion
of a bulge component is not necessary. It has also been shown by
\cite{2006ApJ...641...21R} that a progenitor gas fraction of 30 per
cent results in remnants with parameters in good agreement with the
Fundamental Plane for elliptical galaxies. Using a simple model for
gas accretion onto a central super-massive black hole it has been
argued, based on simulations of binary disk mergers, that gas inflow
regulated by black hole feedback can naturally explain the observed
present day relation between stellar velocity dispersion and black
hole mass for elliptical galaxies and their evolution  with redshift
\cite[(Di Matteo et al.2005)]{2005Natur.433..604D}. Using the same approach we simulated a
large sample of unequal mass binary mergers including
radiative gas cooling, star formation, as well black hole growth and
the associated feedback processes (see Johansson et al. 2007 in prep
and \cite{2005MNRAS.361..776S} for details on the feedback model). We
find that unequal 3:1 mergers follow the same $M_{\rm BH}-\sigma$
relation established for 1:1 mergers, with the scatter primarily
caused by the initial gas mass fraction (see Fig. \ref{naab_fig1}). In addition
we find evidence that also lower ratio mergers of 10:1, as well as
re-simulations of the merger remnants ('E-E' mergers) follow this same
universal correlation between the final black hole mass and the
stellar velocity dispersion. 

Extended models based on mergers in a cosmological context by
\cite{2006ApJS..163....1H} have also been used to explain the
evolution of Quasars and stellar spheroids as a whole. However, recent
studies of dissipative mergers by \cite{2006MNRAS.372..839N}, and,
including star formation and feedback processes, by
\cite{2006ApJ...650..791C} confirm and strengthen
the \cite{2003ApJ...597..893N} conclusion that binary disk mergers
are reasonable progenitors of intermediate mass giant ellipticals but
not for the more massive ellipticals.

\section{The challenges}
\begin{figure}
\begin{center}
 \includegraphics[width=9cm]{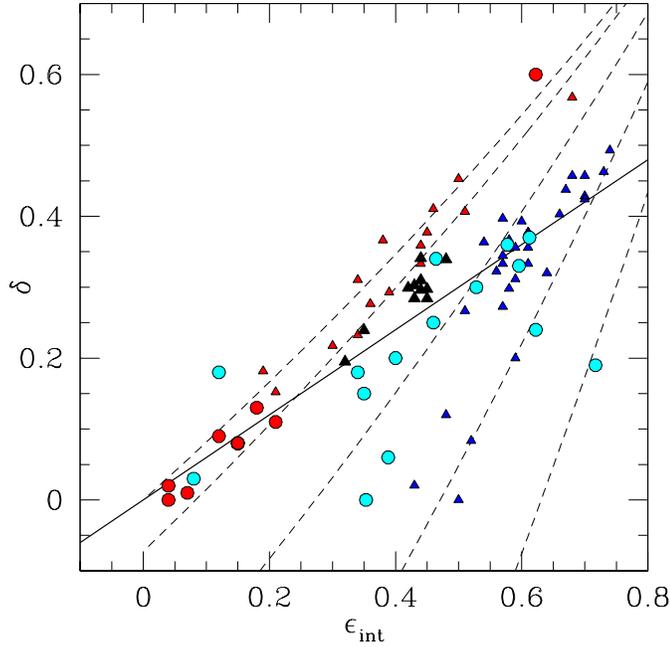}
  \caption{The anisotropy $\delta$ versus the edge-on projected
  ellipticity $\epsilon_{\mathrm{int}}$ is shown for SAURON early-type
  galaxies (slow rotators: red circles, fast rotators: cyan circles)
  and dynamically relaxed merger remnants of disk galaxies mass ratios of 1:1 (red
  triangles) and 3:1 (blue triangles). Mergers between early-type
  galaxies are shown as black triangles. The solid line is the
  correlation derived from modelling given by \cite{2007MNRAS.379..418C}. Dashed lines show
  from left to right the theoretically predicted correlation $\delta(c)$ for
  constant values of $v/\sigma = 0,0.25,0.5,0.75,1$, respectively. No
  simulated binary merger covers the region of the slowly rotating
  SAURON ellipticals, not even early-type mergers.}\label{naab_fig2}
\end{center}
\end{figure}

Still, serious unsolved problems remain as typical disks are less massive
and metal rich than typical ellipticals \cite[(Naab \&
  Ostriker 2006,2007)]{2006MNRAS.366..899N,2007astro.ph..2535N}. It has been shown by \cite{2001ApJ...554..291C}
that the outer kinematics of low mass ellipticals cannot be explained
by unequal mass mergers, a problem not yet solved. Additionally, gas-poor as well as  
gas-rich disk mergers fail to fit the correlation between anisotropy
$\delta$ and ellipticity $\epsilon_{\mathrm{int}}$
inferred  from Schwarzschild modelling of SAURON ellipticals \cite[(Cappellari et
  al. 2007)]{2007MNRAS.379..418C}. In particular, there seems 
to be no way to produce round, isotropic and slowly rotating systems
by binary mergers, not even by mergers of ellipticals (see
Fig. \ref{naab_fig2}). Future investigations, 
including a detailed analysis of simulated merger remnants using orbit
modelling techniques \cite[(Thomas et al. 2007)]{2007arXiv0708.2205T}
will help to understand this discrepancy. It might turn out that
massive ellipticals cannot be made by binary mergers but form in a
more complicated way that can only be investigated using detailed
high resolution cosmological simulations as outlined in 
\cite[Naab et al. (2007)]{2007ApJ...658..710N}.

\begin{discussion}

\end{discussion}

\end{document}